\documentclass{ws-ijmpcs}

\newcommand{\dslash}{\partial \hskip -0.6em /}
\newcommand{\Dslash}{D \hskip -0.7em /}
\newcommand{\zr}[1]{\mbox{\hspace*{#1em}}}
\newcommand{\ID}{\mbox{{\sf 1}\zr{-0.16}\rule{0.04em}{1.55ex}\zr{0.1}}}
\renewcommand{\trimmarks}{} 

\begin{document}

\markboth{H. Weigel, M. Quandt, N. Graham}
{Cosmic Strings Stabilized by Fermion Fluctuations}

%
%

\title{Cosmic Strings Stabilized by Fermion Fluctuations}

\author{Herbert Weigel}

\address{Physics Department, Stellenbosch University,\\
Matieland 7602, South Africa\\
weigel@sun.ac.za}

\author{Markus Quandt}

\address{Institute for Theoretical Physics, T\"ubingen University\\
D--72076 T\"ubingen, Germany\\
quandt@tphys.physik.uni-tuebingen.de}

\author{Noah Graham}

\address{Department of Physics, Middlebury College\\
Middlebury, VT 05753, USA\\
ngraham@middlebury.edu}

\maketitle

\begin{history}
\received{Day Month Year}
\revised{Day Month Year}
\end{history}

\begin{abstract}
We provide a thorough exposition of recent results 
on the quantum stabilization of cosmic strings. Stabilization
occurs through the coupling to a heavy fermion doublet in a
reduced version of the standard model. The study combines the
vacuum polarization energy of fermion zero-point fluctuations
and the binding energy of occupied energy levels, which are of
the same order in a semi-classical expansion. Populating these
bound states assigns a charge to the string. Strings carrying
fermion charge become stable if the Higgs and gauge fields are
coupled to a fermion that is less than twice as heavy as the
top quark. The vacuum remains stable in the model, because
neutral strings are not energetically favored. These findings
suggest that extraordinarily large fermion masses or unrealistic
couplings are not required to bind a cosmic string in the
standard model.

\end{abstract}

\ccode{PACS numbers: 11.27.+d, 03.65.Ge, 14.65.Jk}

\section{Introduction}

It is well--known that the electroweak standard model and many of its
extensions have the potential to support string--like configurations that 
are the particle physics analogs of vortices or  magnetic flux tubes in 
condensed matter physics. Such objects are usually called {\it cosmic strings} 
to distinguish them from the fundamental variables in string theory, and 
also to indicate that they typically stretch over cosmic length scales.
They are also called electroweak strings or
$Z$--strings\cite{Vachaspati:1992fi}\cdash\cite{Nambu:1977ag} 
because the $Z$--component of the electroweak gauge boson acquires the 
structure of the Abelian Nielsen--Olesen vortex\cite{Nielsen:1973cs}. 

Such strings may have emerged copiously in the early universe at interfaces 
between regions of different vacuum expectation values of the Higgs field(s) 
in several stages after the Big Bang until electroweak symmetry breaking was 
reached. If strings are absolutely stable they should have survived and we 
should be able to observe them today. In the electroweak standard model 
topologically stable strings are ruled out, but the absence of this particular 
stabilizing mechanism does not imply that electroweak strings are unstable or 
irrelevant for particle physics. While their direct gravitational
effects are negligible, $Z$--strings can still be relevant for cosmology
at a sub--dominant level\cite{Achucarro:2008fn,Copeland:2009ga}.
Their most interesting consequences originate, however, from their
coupling to the standard model fields. $Z$--strings provide a source
for primordial magnetic fields\cite{Nambu:1977ag}
and they also offer a scenario for baryogenesis with a second order
phase transition\cite{Brandenberger:1992ys,Brandenberger:1994bx}.
In contrast, a strong first order transition as required by the usual bubble
nucleation scenario is unlikely in the electroweak standard model\cite{EWPhase}
without non-standard additions such as supersymmetry or higher--dimensional
operators\cite{Grojean:2004xa}.

These interesting effects are only viable if $Z$--strings are
energetically stabilized by their coupling to the remaining quantum fields.
The most important contributions are expected to come from (heavy) fermions,
since their quantum energy dominates in the limit $N_C \to \infty$, where
$N_C$ is the number of QCD colors or of any other internal degree of freedom.
The Dirac spectrum in typical string backgrounds is deformed to contain either
an exact or near zero mode, so that fermions can substantially lower their
energy by binding to the string. This binding effect can overcome the
classical energy required to form the string background. However, the
remaining spectrum of modes is also deformed and for consistency
its contribution (the vacuum polarization energy) must be taken into account as
well. Heavier fermions are expected to provide more binding since the
energy gain per fermion charge is higher; a similar conclusion can also 
be obtained from decoupling arguments\cite{D'Hoker:1984ph}. 

A number of previous studies have investigated quantum properties of string
configurations. Naculich\cite{Naculich:1995cb} has shown that in the
limit of weak coupling, fermion fluctuations destabilize the string.
The quantum properties of $Z$--strings have also been connected to
non--perturbative anomalies\cite{Klinkhamer:2003hz}.
The emergence or absence of exact neutrino zero modes in a
$Z$--string background and the possible consequences for the
string topology were investigated in Ref. \refcite{Stojkovic}.
A first attempt at a full calculation of the fermionic quantum corrections
to the $Z$--string energy was carried out in Ref.~\refcite{Groves:1999ks}.
Those authors were only able to compare the energies of two
string configurations, rather than comparing a single string
configuration to the vacuum because of limitations arising from the 
non--trivial behavior at spatial infinity. We will discuss a solution
to this obstacle in detail below. 
The fermionic vacuum polarization energy of the Abelian Nielsen--Olesen
vortex has been estimated in Ref. \refcite{Bordag:2003at}
with regularization limited to the subtraction of the divergences
in the heat--kernel expansion. Quantum energies of bosonic fluctuations
in string backgrounds were calculated in Ref. \refcite{Baacke:2008sq}.
Finally, the dynamical fields coupled to the string can also result in
(Abelian or non--Abelian) currents running along the core of the string.
The time evolution of such structured strings was studied in
Ref. \refcite{Lilley:2010av}, where the current was induced by the
coupling to an extra scalar field.

This presentation is based on a number of 
publications\cite{Weigel:2009wi}\cdash\cite{Graham:2011fw} regarding the 
contribution of fermion quantum corrections to the vacuum polarization 
energy of a straight and infinitely long cosmic 
string that appeared over the last year or so. Technical details underlying 
the results given here can be learned from those publications, 
{\it cf.} in particular the appendices of Ref. \refcite{Graham:2011fw}.

\section{Model}

For the current investigation the fermion doublet will be considered 
degenerate so that the introduction of a matrix notation for the Higgs 
field is appropriate. Then the string configuration reads
\begin{equation}
\Phi=vf_H(\rho)
\begin{pmatrix}
{\rm sin}(\xi_1)\, {\rm e}^{-in\varphi} & -i {\rm cos}(\xi_1) \cr
-i {\rm cos}(\xi_1) & {\rm sin}(\xi_1)\, {\rm e}^{in\varphi}
\end{pmatrix}
\label{eq:Higgs}
\end{equation} 
for the Higgs field and
\begin{equation}
\vec{W}=n\,{\rm sin}(\xi_1)\frac{f_G(\rho)}{g\rho}\hat{\varphi}
\begin{pmatrix}
{\rm sin}(\xi_1) & i {\rm cos}(\xi_1)\,{\rm e}^{-in\varphi} \cr
-i {\rm cos}(\xi_1)\,{\rm e}^{in\varphi} & - {\rm sin}(\xi_1)
\end{pmatrix}
\label{eq:gauge}
\end{equation}
for the gauge boson (in temporal gauge). The variables $\rho$ and 
$\varphi$ are polar coordinates in the plane perpendicular to the string, 
while the Higgs vacuum expectation value~$v$ and the gauge coupling 
constant $g$ are model parameters. The string configuration involves profile 
functions $f_H$ and $f_G$ which are the analogs of the Nielsen--Olesen vortex 
profiles with boundary conditions
\begin{equation}
\begin{array}{ll}
\rho\,\longrightarrow\, 0:\qquad &
f_G\,,\,f_H\, \longrightarrow\, 0 \cr
\rho\,\longrightarrow\, \infty:\qquad &
f_G\,,\,f_H\, \longrightarrow\, 1\,.
\end{array}
\label{eq:bc}
\end{equation}
The integer $n$ is the winding of the string, for which we will typically 
take $n=1$ in numerical calculations. Finally, the (variational) parameter 
$\xi_1$ measures the relative strength of the Higgs and gauge boson components 
of the string.

We will consider a modified version of the electroweak standard model which 
only has a $SU(2)_L$ gauge symmetry. This modification is equivalent to a 
vanishing Weinberg angle. Then the (classical) boson part of the Lagrangian reads
\begin{equation}
\mathcal{L}_{\rm bos}=-\frac{1}{2} {\rm tr}\left(G^{\mu\nu}G_{\mu\nu}\right) 
+\frac{1}{2} {\rm tr} \left(D^{\mu}\Phi \right)^{\dag} D_{\mu}\Phi
- \frac{\lambda}{2} {\rm tr} \left(\Phi^{\dag} \Phi - v^2 \right)^2 \,,
\label{eq:Lboson}
\end{equation}
with the covariant derivative $D_\mu = \partial_\mu - i \,g W_\mu$ 
and the $SU(2)$ field  strength tensor 
\begin{equation}
G_{\mu\nu}=\partial_\mu\,W_\nu-\partial_\nu W_\mu-ig\left[W_\mu,W_\nu\right]\,.
\label{fieldtensor}
\end{equation}
The boson masses are determined from $g$ and $v$ and the Higgs self--coupling 
$\lambda$ as $m_{\rm W}=gv/\sqrt{2}$ and $m_{\rm H}=2v\,\sqrt{\lambda}$ for the 
gauge and Higgs bosons, respectively. The interaction with the fermion doublet 
is described by the Lagrangian
\begin{equation}
\mathcal{L}_{\rm fer}=i\overline{\Psi}
\left(P_L \Dslash  + P_R \dslash \right) \Psi
-f\,\overline{\Psi}\left(\Phi P_R+\Phi^\dagger P_L\right)\Psi
\quad {\rm with} \quad
P_{R,L}=\frac{1}{2}\left(\ID\pm\gamma_5\right)\,.
\label{eq:Lfermion}
\end{equation}
Upon spontaneous symmetry breaking the Yukawa coupling $f$ induces 
a fermion mass $m=vf$. The similarity with the standard model of
particle physics suggests the model parameters
\begin{equation}
g=0.72\,,\quad
v=177\,{\rm GeV}\,,\quad
m_{\rm H}= 140\,{\rm GeV}\,,\quad
f=0.99\,,
\label{eq:parameters}
\end{equation}
where we have taken the fermion doublet to have the mass of the top quark. 
In our numerical search for a stable string, we will also study other model 
parameters, {\it cf.} section \ref{sec:charged}.

The classical energy per unit length of the string is solely governed
by $\mathcal{L}_{\rm bos}$:
\begin{eqnarray}
\frac{E_{\rm cl}}{m^2}&=&2\pi\int_0^\infty \rho\, d\rho\,\Biggl\{
n^2\sin^2 \xi_1\,\biggl[\frac{2}{g^2}
\left(\frac{f_G^\prime}{\rho}\right)^2
+\frac{f_H^2}{f^2\rho^2}\,\left(1-f_G\right)^2\biggr]
\cr\cr &&\hspace{3cm}
+\frac{f_H^{\prime2}}{f^2}
+\frac{\mu_H^2}{4f^2}\left(1-f_H^2\right)^2\Biggr\}\,,
\label{eq:Ecl}
\end{eqnarray}
where the dimensionless radial integration variable is related to the 
physical radius by $\rho_{\rm phys}=\rho/m$ and $\mu_{H}\equiv m_{\rm H}/m$. 

The central object of the present investigation is the fermion 
contribution to  the energy. It will obtained from the solutions to the Dirac 
equation\footnote{We use the standard representation with $\gamma_0={\rm diag}(1,1,-1,-1)$.} 
in the two--dimensional plane perpendicular to the string
\begin{equation}
H\Psi_n=\omega_n \Psi_n 
\quad {\rm with} \quad
H=-i\begin{pmatrix}0 & \vec{\sigma}\cdot\hat{\rho} \cr
\vec{\sigma}\cdot\hat{\rho} & 0\end{pmatrix} \partial_\rho
-\frac{i}{\rho}\begin{pmatrix}0 & \vec{\sigma}\cdot\hat{\varphi} \cr
\vec{\sigma}\cdot\hat{\varphi} & 0\end{pmatrix} \partial_\varphi
+H_{\rm int}\,,
\label{eq:Dirac}
\end{equation}
where the single particle Hamiltonian, $H$ is extracted from $\mathcal{L}_{\rm fer}$.
The interaction part, $H_{\rm int}$ depends on the chosen gauge. A specific 
choice of gauge, which accommodates the subtleties at large $\rho$, will be discussed 
in  Sec.~\ref{subsec:Gauge}. The spectrum of the Dirac operator consists of bound 
state solutions with discrete eigenvalues $\omega_n=\epsilon_j$ and continuous 
scattering solutions whose eigenvalues $\omega$ are labeled by momentum $k$, 
i.e.~$\omega=\sqrt{k^2+m^2}$. Finally, the Dirac Hamiltonian for the string background 
anti--commutes with $\alpha_3=\gamma^0\gamma^3$. Thus the spectrum is charge 
conjugation invariant and it suffices to consider the non--negative eigenvalues.

\section{Formalism}

The vacuum polarization energy is the renormalized sum of the changes of the 
zero--point energies of fermions in the background of a static configuration
\begin{equation}
E_{\rm vac}=-\frac{\hbar}{2}\sum_n\left(\omega_n-\omega_n^{(0)}\right)
\Bigg|_{\rm ren}
=-\frac{\hbar}{2}\sum_j \epsilon_j -
\hbar \int_0^\infty dk\, \omega_k\,\Delta\,\rho_{\rm ren}(k)\,.
\label{eq:def}
\end{equation}
Here $\omega_n$ are the energy eigenvalues in the presence of the 
string as obtained from the Dirac equation (\ref{eq:Dirac}), and 
$\omega_n^{(0)}$ are their free counterparts. The overall sign has 
been chosen to describe fermionic vacuum fluctuations; for 
bosons it must be altered. In the second part of equation~(\ref{eq:def}) 
the changes of the single particle energies are re--written as 
the contribution from distinct bound states ($\epsilon_j$) and the 
(renormalized) change of the density of scattering states, 
$\Delta\,\rho_{\rm ren}(k)$. These states obey the standard dispersion relation
$\omega_k=\sqrt{k^2+m^2}$ with the mass $m$ of the quantum fluctuations. 
In the above, the factor $\hbar$ has been made explicit to stress that 
$E_{\rm vac}$ is a quantum effect. As in the previous section natural 
units ($\hbar=1$ and $c=1$) will be adopted in the remainder of this paper.

The computation of the fermion contribution to the vacuum polarization 
energy of a cosmic string proceeds in three stages. First, spectral 
methods are employed to express this energy in form of scattering 
data\cite{Graham:2009zz}. The equivalence of terms in the Born series 
and Feynman diagrams makes it possible to impose standard 
renormalization conditions. In a second step this approach is extended 
to accommodate configurations that are translationally invariant in 
one or more spatial directions, using the so--called interface 
formalism\cite{Graham:2001dy}. Third, we have to cope with the fact that 
the cosmic string configuration has a non--trivial structure at spatial 
infinity. Because the fields approach a pure gauge rather than zero, the 
direct application of spectral methods is impossible. Rather a particular 
subset of gauges has to be adopted to define a well--behaved scattering 
problem\cite{Weigel:2010pf}.

\subsection{Spectral Methods}

The basic idea of the spectral approach is to express 
$\Delta\,\rho_{\rm ren}(k)$ as the momentum derivative of the phase shifts 
of the scattering states\cite{Krein:1953}. Then
\begin{equation}
E_{\rm vac}=-\frac{1}{2}\sum_j \epsilon_j -
\sum_{\ell} D_\ell \int_0^\infty \frac{dk}{2\pi}\, \omega_k\, 
\frac{d}{dk}\left[\delta_\ell(k)\right]_{\rm ren}\,.
\label{eq:attempt1}
\end{equation}
It is implicitly assumed 
that the system allows for a partial wave decomposition with degeneracy 
factor $D_\ell$ for the partial wave of (generalized) angular momentum~$\ell$. 
The major concern in eq.~(\ref{eq:attempt1}) is renormalization. 
As it stands, in three spatial dimensions the vacuum polarization energy, 
eq.~(\ref{eq:attempt1}), is quadratically divergent in the ultra--violet. 
For large momenta the Born series adequately represents the phase 
shift\footnote{For small momenta it does generally not converge towards the 
exact phase shift, in particular when bound states are present.}. We 
therefore consider 
\begin{equation}
E^{(N)}=-\frac{1}{2}\sum_j \epsilon_j -
\sum_{\ell} D_\ell \int_0^\infty \frac{dk}{2\pi}\, \omega_k\,
\frac{d}{dk}\left[\delta_\ell(k)\right]_N\,,
\label{eq:attempt2}
\end{equation}
where the subscript on the phase shift indicates the subtraction of 
the $N$ leading terms of the Born series to the phase shift: 
$\left[\delta_\ell\right]_N=\delta_\ell-\delta_\ell^{(1)}-\delta_\ell^{(1)}
-\ldots-\delta_\ell^{(N)}$.
Choosing $N$ sufficiently large renders $E^{(N)}$ finite. Technically
the Born series is an expansion in powers of $H_{\rm int}$. The contribution
of any such power to the vacuum polarization energy can be associated with
a Feynman diagram

\begin{equation}
\sum_{\ell} D_\ell
\int_0^\infty \frac{dk}{2\pi}\, \omega_k\, \frac{d}{dk}
\delta_\ell^{(n)}(k)\quad \mbox{\LARGE $\sim$}\qquad
\mbox{\parbox[l]{3.5cm}{\vspace{-0.4cm}
\psfig{file=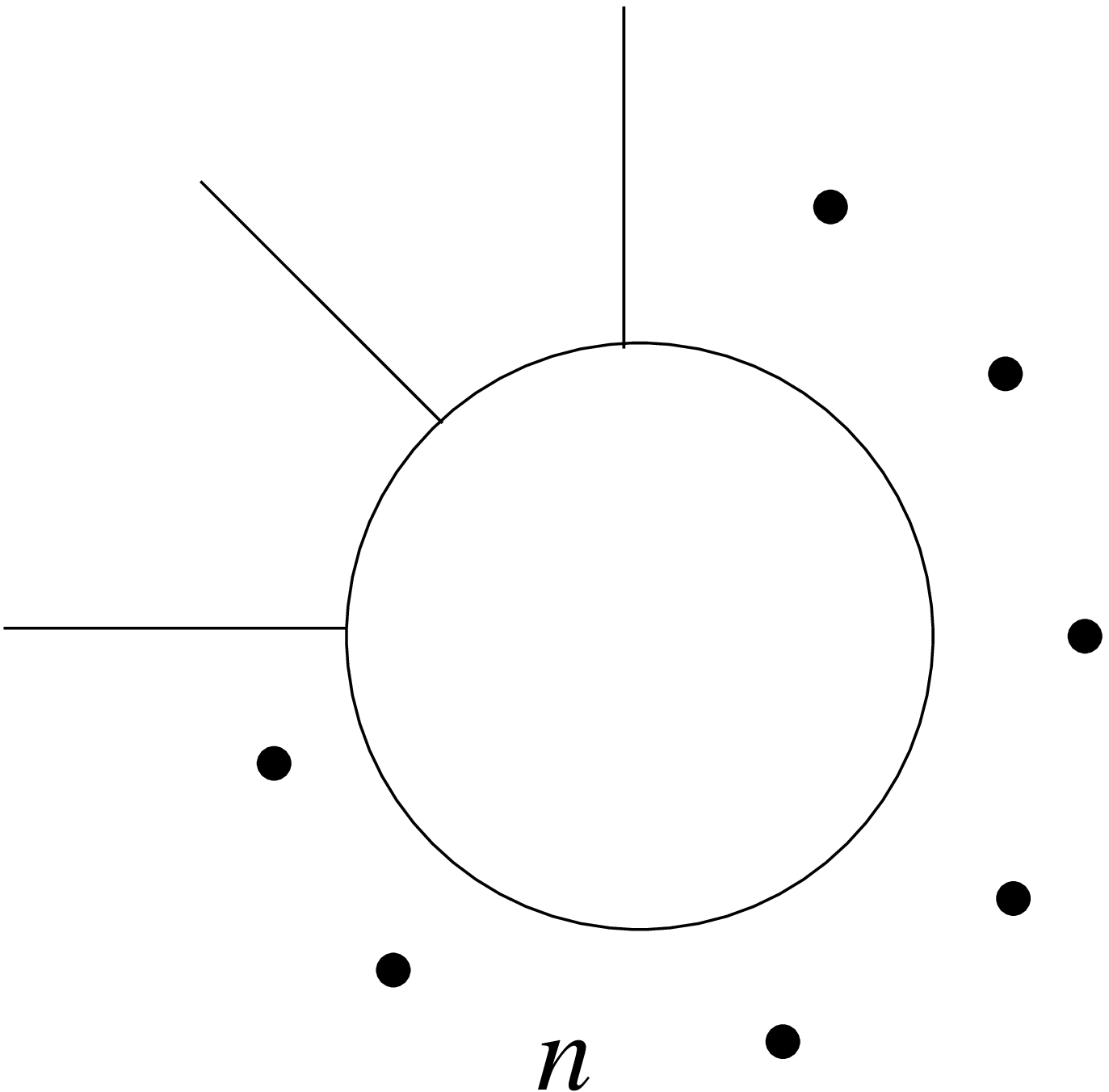,height=2cm,width=3cm}}}\,,
\label{eq:EbornFD}
\end{equation}
where the loop corresponds to the quantum fluctuation and the external lines
represent insertions of the background field, {\it i.e.} the interaction of
the cosmic string with the fermions. Therefore, the artificial subtraction
in equation~(\ref{eq:attempt2}) is the sum over all Feynman diagrams with 
up to $N$ insertions of the background interaction: $E^{(N)}_{\rm FD}$. These 
diagrams can be computed with standard techniques, in particular dimensional 
regularization can be implemented to handle the ultra--violet divergences. 

The most important advantage of shuffling the divergences into Feynman 
diagrams is, however, that they can be straightforwardly combined with the 
counterterm contribution to the energy, $E_{\rm CT}$. The latter is found by 
substituting the background configuration into the counterterm Lagrangian in 
just the same way as the classical energy, eq.~(\ref{eq:Ecl}) is obtained
from $\mathcal{L}_{bos}$. In any (multiplicatively) renormalizable theory the 
counterterm Lagrangian has the same structure as the classical one and a 
suitable choice of its coupling constants cancels all ultra--violet divergences 
completely.  Then $E^{(N)}_{\rm FD}+E_{\rm CT}$ is free of any divergences. The finite 
pieces of the counterterm coupling constants will be uniquely determined from 
appropriate conditions describing properties of the particles that are 
associated with the fields. This procedure will be briefly discussed in 
section.~\ref{subsec:Onshell}.  In total, the sum
\begin{equation}
E_{\rm vac}=E^{(N)}+E^{(N)}_{\rm FD}+E_{\rm CT}
\label{eq:Efinite}
\end{equation}
gives an unambiguous result for the vacuum polarization energy once
the renormalization conditions are fixed. A first principle derivation 
of this result based on a quantum field theoretic formulation of the 
energy momentum momentum tensor rather and the analytic properties of 
the Greens function for scattering boundary conditions is presented 
in Ref.~\refcite{Graham:2002xq}. 

\subsection{Interface Formalism}

Equation~(\ref{eq:Efinite}) is perfectly suited for a computation of the vacuum 
polarization energy for static configurations that allow for a full decomposition
into partial waves. However, this is not the case for the cosmic string 
which is translationally invariant along the $\hat{z}$--axis in coordinate
space. In this scenario the wave--function of the quantum fluctuation 
factorizes into 
\begin{equation}
\Psi(\vec{x},t)\sim {\rm e}^{-i\omega t} \, {\rm e}^{ipz}\,
\psi_{k}(\vec{\rho})
\label{eq:Inface1}
\end{equation}
so that we have a scattering problem in the plane perpendicular to the string, 
and the total dispersion relation $\omega=\sqrt{p^2+k^2+m^2}=\sqrt{p^2+\omega_k^2}$.
This implies the replacements
$\epsilon_j\to\sqrt{p^2+\epsilon_j^2}$ and 
$\omega_k\to\sqrt{p^2+\omega_k^2}$ in equation~(\ref{eq:attempt2}) so that an 
integration over $p$ (with measure $dp/2\pi$) yields the vacuum polarization
energy per unit length of the string. In doing so an immediate obstacle 
arises. The scattering data do not depend on the momentum $p$ along the 
symmetry axis. Hence for any~$N$, this $p$--integral will not be finite. A 
careful analysis treats the $p$--integral in dimensional 
regularization\cite{Graham:2001dy}
\begin{equation}
E^{(N)}\sim \frac{\Gamma(-\frac{1+d}{2})}{2(4\pi)^{\frac{d+1}{2}}}
\sum_\ell D_\ell
\Bigg\{\sum_j\left(\epsilon_j\right)^{\frac{d+1}{2}} 
+ \int_0^\infty \frac{dk}{\pi} \left(k^2+m^2\right)^{\frac{d+1}{2}}\,
\frac{d}{dk}\left[\delta_\ell(k)\right]_{N}\Big]\Bigg\}\,,
\label{eq:Inface2}
\end{equation}
where $d$ is the analytic dimension of the subspace in which the configuration
is translationally invariant. The divergence now manifests itself via the 
singularity of the $\Gamma$--function coefficient as $d\to1$. Due to sum rules 
for scattering data\cite{Puff:1975zz}, which represent generalizations of 
Levinson's theorem, the expression in curly brackets in eq.~(\ref{eq:Inface2})
vanishes as $d\to1$. Hence this limit can indeed be taken\cite{Graham:2001dy}. 
Finally, the fermion spectrum in the string background is charge conjugation 
invariant. Thus $E^{(N)}$ 
is twice its contribution from the non--negative part of the spectrum,
\begin{eqnarray}
E^{(N)}&=&\frac{1}{4\pi}\sum_\ell D_\ell \Bigg\{
\int_0^\infty \frac{dk}{\pi}
\left[\omega_k^2\, {\rm ln}\left(\frac{\omega_k^2}{\mu_r^2}\right)-k^2\right]\,
\frac{d}{dk}\left[\delta_\ell(k)\right]_N \cr
&& \hspace{4cm}
+\sum_j\left[\epsilon_j^2\, {\rm ln}\frac{\epsilon_j^2}{\mu_r^2}
-\epsilon_j^2+m^2\right]\Bigg\}\,.
\label{eq:Ereg}
\end{eqnarray}
Here $\mu_r$ is an arbitrary renormalization scale that 
has no effect on $E^{(N)}$ by exactly the same sum rules. 
The expression, eq.~(\ref{eq:Ereg}), for $E^{(N)}$ replaces the analog in 
equation~(\ref{eq:Efinite}). Note that the function multiplying the
(Born subtracted) phase shift is of higher power in $k$ than its counterpart
before integrating over the momentum $p$ conjugate to the coordinate of 
translational invariance. Hence $N$ must be increased if directions are 
added in which the configuration is translationally invariant. This, of course,
merely reflects the fact that ultra--violet divergences turn more severe
in higher dimensions.

We stress that the expressions obtained so far for the vacuum polarization
energy strongly rely on the analytic properties of the scattering data. 
Furthermore, in numerical calculations, we mostly consider them for 
purely imaginary momenta. This has (at least) two advantages: (i) the
oscillating phase shifts turn into exponentially decaying (logarithms
of the) Jost functions, and (ii) the momentum integral and the sum over 
angular momenta may be exchanged. While (i) drastically improves numerical 
stability, (ii) significantly simplifies the treatment
of the logarithmic divergences that emerge at third and fourth order
of the Born and Feynman expansions. These technical details are discussed 
at length in Ref.~\refcite{Graham:2011fw} and briefly addressed in 
section~\ref{sec_fake}.

\subsection{Choice of Gauge}

We have now established a formalism for computing the vacuum polarization
energy of background fields in the string geometry. However, we still have 
the problem that the string does not induce a well--behaved scattering 
problem because of its non--trivial structure at spatial infinity. Though 
gauge invariant combinations of the Higgs and gauge bosons are trivial at 
spatial infinity, the individual terms in the Born and Feynman series are not
gauge invariant and therefore ill--defined. This ambiguity appears because 
the Dirac Hamiltonian that is obtained by straightforward substitution
of the field configuration, eqs.~(\ref{eq:Higgs}) and~(\ref{eq:gauge})
does not turn into the free Dirac Hamiltonian as $\rho\to\infty$, but
instead becomes
$H\to U^\dagger(\varphi) H_{\rm free} U(\varphi)$. This local gauge 
transformation acts only on the left--handed fermions,
$U(\varphi)=P_L {\rm exp}\left(i\hat{n}\cdot\vec{\tau}\,\xi_1\right)+P_R$
with $\hat{n}=\left({\rm cos}(n\varphi),-{\rm sin}(n\varphi), 0\right)$\,.
Unfortunately, the gauge transformation $H \to U(\varphi)HU^\dagger(\varphi)$ 
does not solve the problem for all $\rho\in[0,\infty]$: Although it would 
generate vanishing interactions at infinity, it also induces a $1/\rho^2$ 
potential at the center of the string, $\rho \to 0$. This might still yield 
well--defined phase shifts, but the conditions underlying the analyticity of 
the scattering data are certainly violated by this singular behavior. As argued 
at the end of the previous section, analyticity is central for numerical 
feasibility of our approach. As a solution, we can define a radially 
extended gauge transformation
\begin{equation}
U(\rho,\varphi)=P_L {\rm exp}\left(i\hat{n}\cdot\vec{\tau}\,\xi(\rho)\right)+P_R\,.
\label{eq:radialGT}
\end{equation}
This transformation fixes the gauge and in equation~(\ref{eq:Dirac}) it 
yields the interaction term
\label{subsec:Gauge}
\begin{eqnarray}
H_{\rm int}&=&
m f_H\left[{\rm cos}(\Delta)\begin{pmatrix} \ID & 0 \cr 0 &-\ID\end{pmatrix}
+i\,{\rm sin}(\Delta)\begin{pmatrix}0 & \ID \cr -\ID & 0\end{pmatrix}
\vec{n}\cdot\vec{\tau}\right]
+\frac{1}{2}\frac{\partial \xi}{\partial \rho}
\begin{pmatrix}-\vec{\sigma}\cdot\hat{\rho}
& \vec{\sigma}\cdot\hat{\rho} \cr
\vec{\sigma}\cdot\hat{\rho}
& -\vec{\sigma}\cdot\hat{\rho}\end{pmatrix}\vec{n}\cdot\vec{\tau}
\cr\cr 
&& \hspace{0.5cm}
+\frac{n}{2\rho}\, \begin{pmatrix}-\vec{\sigma}\cdot\hat{\varphi}
& \vec{\sigma}\cdot\hat{\varphi} \cr
\vec{\sigma}\cdot\hat{\varphi}
& -\vec{\sigma}\cdot\hat{\varphi}\end{pmatrix}
\Big[f_G\,{\rm sin}(\Delta)I_G(\Delta)+(f_G-1)\,{\rm sin}(\xi)I_G(-\xi)\Big]\,.
\label{eq:DiracInt}
\end{eqnarray}
The new gauge function $\xi(\rho)$ is hidden in the difference
$\Delta(\rho)  \equiv \xi_1 - \xi(\rho)$ which appears both explicitly and
as the argument of the space--dependent weak isospin matrix
\begin{equation}
I_G(x)=\begin{pmatrix}-{\rm sin}(x) & -i\,{\rm cos}(x)\,{\rm e}^{in\varphi} \\[2mm]
i\,{\rm cos}(x)\,{\rm e}^{-in\varphi} & {\rm sin}(x) \end{pmatrix} \,.
\label{defIG}
\end{equation}
Imposing the boundary conditions $\xi(0)=0$ and $\xi(\infty)=\xi_1$ for the
new gauge function $\xi(\rho)$ defines a well--behaved scattering problem.
Otherwise, the specific form of $\xi(\rho)$ is irrelevant. 
This property allows us to verify our numerical results by modifying 
its shape while keeping the boundary conditions fixed.

All explicit matrices in eq.~(\ref{eq:DiracInt}) act in spinor space.
Together with the boundary conditions, eq.~(\ref{eq:bc}) a well--behaved 
scattering problem is obtained. With this choice of gauge a scattering matrix 
and, more generally, a Jost function can be straightforwardly computed. 
Moreover, the Born series to these scattering data can be constructed 
simply by iterating $H_{\rm int}$.

Note that the gauge transformation is single--valued at spatial infinity, 
$U(\infty,\varphi)=U(\infty,\varphi+2\pi)$. In this respect it differs from
the analogous problem of fractional fluxes in QED. In that case a similar choice 
of gauge is hence not a remedy; rather the calculation of the vacuum polarization 
energy requires the introduction of a return flux to arrive at a well--behaved 
scattering problem\cite{Graham:2004jb}.  The return-flux approach can also be 
used for the present calculation, but it is much more laborious 
numerically\cite{Weigel:2009wi,Weigel:2010pf}.

\subsection{Fake Boson Field}
\label{sec_fake}

The idea of utilizing a fake boson field to simplify the treatment of
higher order divergences was first implemented in Ref.~\refcite{Farhi:2001kh}.
As mentioned above, the continuation to imaginary momenta $k\to it$ and 
$\delta_\ell(k)\to\nu(t)$, where $\nu$ is the logarithm of the Jost function, 
allows the exchange of the momentum integral with the angular momentum sum. 
Then the third and fourth order contribution from the Born series produce 
logarithmic divergences. These divergences are similar to the ones found in 
the second order vacuum polarization energy of a boson field fluctuating about 
a scalar potential. Matching its strength appropriately allows to replace
\begin{equation}
\sum_\ell D_\ell \left[\frac{d}{dt}\nu_\ell(t)\right]_N\,\longrightarrow\,
\frac{d}{dt}\left[\sum_\ell D_\ell\left(
\nu_\ell(t)-\nu^{(1)}_\ell(t)-\nu^{(2)}_\ell(t)\right)
-\sum_{\ell}\bar{D}_{\ell}\bar{\nu}^{(2)}_\ell(t)\right]
\label{eq:FakeBos}
\end{equation}
under the integral in eq.~(\ref{eq:Ereg}). The quantity $E^{(N)}$ with this 
replacement will be called~$E_\delta$. The over--bared quantities refer to the 
bosonic scattering data. The replacement eq.~(\ref{eq:FakeBos}) must, of course, 
be accompanied by the boson Feynman diagram~$E_{\rm B}$ so that the total 
vacuum polarization energy becomes
\begin{equation}
E_{\rm vac}=E_\delta+E_{\rm FD}^{\rm ren.}+E_{\rm B}^{\rm ren.}\,,
\label{eq:Master}
\end{equation}
where the superscript indicates the inclusion of the counterterm 
contributions. Each of the three terms on the right hand side of 
equation~(\ref{eq:Master}) is ultra--violet finite by itself.
The advantage of eq.~(\ref{eq:Master}) and the replacement 
eq.~(\ref{eq:FakeBos}) is now obvious: Instead of fermionic 
contributions up to order $N=4$, we only need to compute second 
order fermionic and bosonic Feynman diagrams and terms in the 
corresponding Born series.

\section{Numerical Results for the Vacuum Polarization Energy}

Numerical results as well as dimensionfull parameters are measured
in appropriate units of the (perturbative) fermion mass $m$.

We focus on the contribution of fermion fluctuations to the vacuum 
polarization energy because it dominates the boson counterpart by
a factor proportional to the number of internal degrees of freedom,
{\it e.g.} $N_C$, the number of colors. In this scenario we are now 
prepared to compute the vacuum polarization energy of a prescribed 
string configuration.

\subsection{Variational Ans\"atze}

Despite of the simplification in eq.~(\ref{eq:Master}), the numerical 
computation is still expensive. The scattering data are extracted from a 
multi--channel problem and for the final result to be reliable a huge 
number of partial wave must be included.  This numerical effort restricts 
the number of variational parameters that can be used to characterize 
the profile functions. We have already introduced the strength parameter 
$\xi_1$. In addition, we introduce three scale parameters $w_H$, $w_W$ 
and $w_\xi$ via the ans\"atze
\begin{equation}
f_H(\rho)=1-{\rm e}^{-\frac{\rho}{w_H}}\,,\quad
f_G(\rho)=1-{\rm e}^{-\left(\frac{\rho}{w_G}\right)^2}\,,\quad
\xi(\rho)=\xi_1\left[1-
{\rm e}^{-\left(\frac{\rho}{w_\xi}\right)^2}\right]\,.
\label{eq:Ansaetze}
\end{equation}
The scale $w_\xi$ parameterizes the shape of the gauge profile. As explained 
above, this shape and thus $w_\xi$ should not be observable. 
The other specifics of the profiles are chosen to keep $E_{\rm cl}$ regular.

We have also considered an exponential parameterization for the gauge
field 
\begin{equation}
f_G(\rho)=1-\left(1+\frac{\rho}{w_G}\right)
{\rm exp}\left(-\frac{\rho}{w_G}\right)\,,
\label{eq:ExpAnsatz}
\end{equation}
which yields a slightly better agreement with the original Nielsen--Olesen
profiles that minimize $E_{\rm cl}$ for $\xi_1=\pi/2$. No significant
difference in $E_{\rm vac}$ was found between these ans\"atze.

\subsection{Gauge Invariance}

We check gauge invariance by varying the shape of the gauge profile, 
$\xi(\rho)$. A typical result is shown in table \ref{tab:Res1}.
\begin{table}[pb]
\tbl{Numerical results for the various contributions~(\ref{eq:Master}) to 
the fermion vacuum polarization energy in the minimal subtraction scheme.}
{\begin{tabular}{l|lll|l}
$w_\xi$ &~ $E_{\rm FD}^{\rm ren.}$ & $E_\delta$ &
$E_{\rm B}^{\rm ren.}$ & $E_{\rm vac}$ \cr
\hline
0.5 &~ -0.2515 & 0.3489 & 0.0046 ~& 0.1020 \cr
1.0 &~ -0.0655 & 0.1606 & 0.0032 ~& 0.0983 \cr
2.0 &~ -0.0358 & 0.1294 & 0.0038 ~& 0.0974 \cr
3.0 &~ -0.0320 & 0.1235 & 0.0056 ~& 0.0971 \cr
4.0 &~ -0.0302 & 0.1193 & 0.0080 ~& 0.0971
\end{tabular}}
\label{tab:Res1}
\end{table}
As expected, the individual contributions to $E_{\rm vac}$ depend strongly on 
$w_\xi$. However, these changes essentially compensate each other. Numerically 
the most cumbersome part of the calculation is $E_\delta$. From various 
numerical considerations (change of extrapolation scheme for partial wave sum, 
modification of momentum integration grid, etc.) its numerical accuracy is 
estimated to be at the 1\% level. Within that range $E_{\rm vac}$ 
is independent of $w_\xi$, thus verifying gauge invariance.

The above results are obtained in the $\overline{\rm MS}$ renormalization 
scheme, which essentially omits the non--divergent parts of the Feynman diagrams. 
Any other scheme merely differs by manifestly gauge invariant (finite) 
counterterms.

\subsection{On--Shell Renormalization}
\label{subsec:Onshell}

\begin{figure}[pb]
\centerline{\psfig{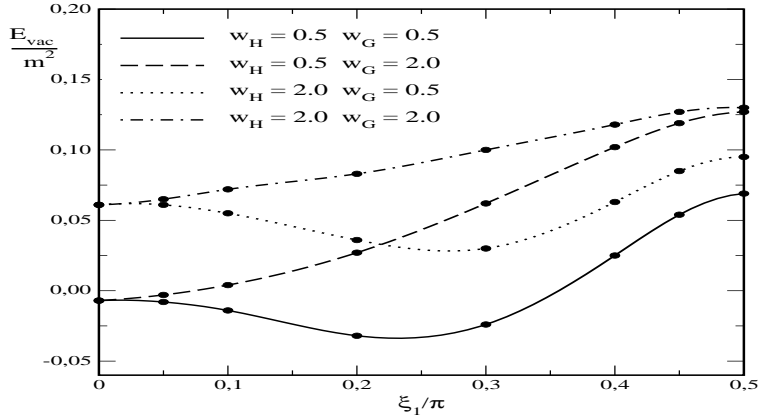}}
\vspace*{8pt}
\caption{Fermion vacuum polarization energy in the on--shell
renormalization scheme. \label{fig_onshell}}
\end{figure}

With the above mentioned choice of units, the dependence of $E_{\rm vac}$
on the model parameters factorizes in the $\overline{\rm MS}$ scheme 
which simplifies the computation because this dependence can easily be
traced from $E_{\rm cl}$. However, for physically meaningful results
we need to impose renormalization conditions that correspond to a particle 
interpretation, inducing a mild parameter dependence in $E_{\rm CT}$. 
To be specific we consider the so--called {\it on--shell} scheme in which 
the coefficients of the four allowed counterterms are determined such that
\begin{itemize}
\item[$\bullet$]
the tadpole graph vanishes
\item[$\bullet$]
the Higgs mass remains unchanged
\item[$\bullet$]
the normalization of Higgs particle remains unchanged
\item[$\bullet$]
and the normalization of vector meson remains unchanged
\end{itemize}
in the presence of fermionic quantum corrections. 
Note that the vector meson mass~$M_W$ is not fixed by these conditions 
and thus will be a prediction that includes quantum corrections. Hence we 
tune the gauge coupling to reproduce the physical value for~$M_W$. Typical 
results for the vacuum polarization energy per unit length of the string are 
shown in figure \ref{fig_onshell}, as functions of the variational parameters.
Except for narrow string configurations dominated by the Higgs field, the vacuum 
polarization energy turns out to be positive. Therefore, fermionic quantum
corrections do not provide any sensible binding and no stable uncharged string 
is found for the physically motivated parameters, eq.~(\ref{eq:parameters}),
for which $E_{\rm cl}$ dominates the total energy. Yet, $E_{\rm cl}$
deceases quickly with increasing Yukawa coupling $f$ and some stability is indeed
seen for large $f$ and narrow strings. Unfortunately, in this regime 
the restriction to one fermion loop in the vacuum polarization energy is unreliable 
because of the occurrence of the Landau ghost\cite{Ripka:1987ne}.

\section{Charged Strings}
\label{sec:charged}

\begin{figure}[pb]
\centerline{\hspace{1.0cm}\psfig{file=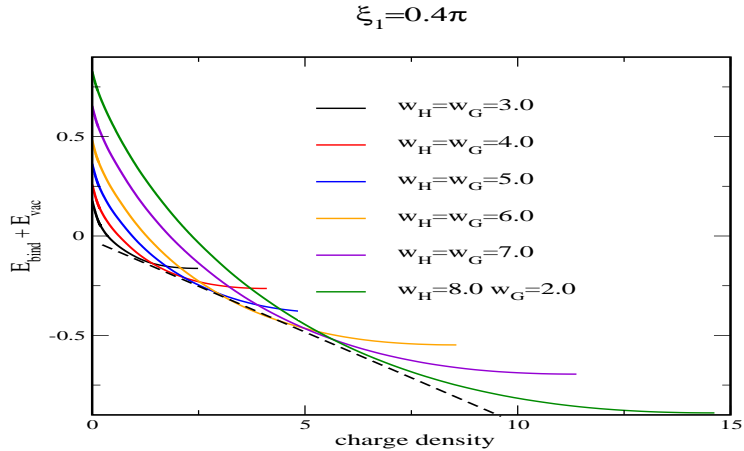,width=11.0cm,height=6.5cm}}
\vspace*{8pt}
\caption{(Color online) Fermion contribution to the energy of a 
charged string. \label{fig_ebline}}
\end{figure}

Cosmic strings induce many fermionic bound state levels (whose energies are
denoted by $\epsilon_j$) for the two--dimensional scattering problem. For 
$\xi_1=\pi/2$ there even exists an exact zero mode\cite{Naculich:1995cb}. In 
the three dimensional problem these bound states acquire a longitudinal momentum 
for the motion along the symmetry axis and their energies become
\begin{equation}
E_i(p_n)=\sqrt{\epsilon_i^2+p_n^2}
\qquad {\rm with} \qquad
p_n=\frac{n\pi}{L}\,.
\label{eq:E3d}
\end{equation}
Here $L$ is the length of the string. In leading order of the $L\to\infty$ limit
the sum over the discrete longitudinal momentum turns into a continuum integral, 
$\sum_n \, \longrightarrow\, \frac{L}{\pi}\int dp$. To minimize the bound state 
contribution a chemical potential $\mu$ with ${\rm max}(|\epsilon_j|)\le\mu\le m$ 
is introduced and all levels with $E_i(p)\le \mu$ are populated. This procedure 
defines a Fermi momentum for each level, $p^F_i(\mu)=\sqrt{\mu^2-\epsilon_i^2}$ 
which enters the total charge per unit length of the string 
\begin{equation}
Q(\mu)=\sum_i \frac{p^F_i(\mu)}{\pi}\,.
\label{eq:TotalQ}
\end{equation}
This relation can be inverted to give $\mu=\mu(Q)$ and thus $p^F_i=p^F_i(Q)$.
From this the binding energy (per unit length) for a prescribed charge
\begin{equation}
E_{\rm bind}(Q)=\frac{1}{\pi}\sum_i\int_0^{p^F_i(Q)}dp \,
\left[\sqrt{\epsilon_i^2+p^2}-m\right]
\label{eq:Ebind}
\end{equation}
is computed relative to an equal number of free fermions that have 
energy~$m$ each.  Figure~\ref{fig_ebline} shows the fermion contribution to the
binding energy, $E_{\rm vac}+E_{\rm bind}(Q)$. For a given configuration
the graph terminates at the point when all available bound state levels
are occupied. For small charges narrow strings are favorable while 
the binding energy of strings with larger widths decreases more quickly
as $Q$ increases.
Surprisingly, the envelope along which $E_{\rm vac}+E_{\rm bind}$ is minimal 
forms a straight line.  Extrapolating this line to~~$Q=0$~~indicates that
the fermion vacuum polarization energy should (approximately) vanish. 
This extrapolation circumvents the Landau ghost problem. 

\begin{figure}[pb]
\vspace*{2pt}
\centerline{\psfig{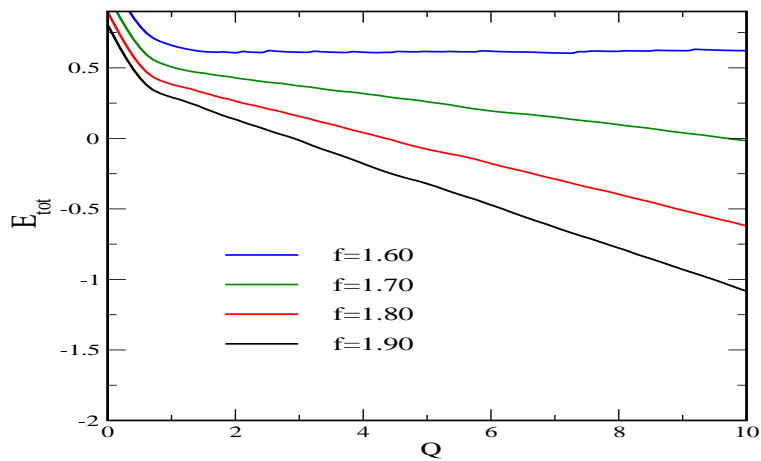}}
\vspace*{6pt}
\caption{(Color online) Total energy of the charged string.
\label{fig_etot1}}
\end{figure}

To finally decide on dynamical stability, the classical energy must be included. 
To this end we scan through several hundred configurations characterized by 
the variational ans\"atze, eq.~(\ref{eq:Ansaetze}). We label them by 
$s=1,2,\dots$ and compute their total binding energy 
\begin{equation}
E^{(s)}_{\rm tot}(Q)=E^{(s)}_{\rm cl}
+N_C\left[E^{(s)}_{\rm vac}+E^{(s)}_{\rm bind}(Q)\right]
\label{eq:Ebindform}
\end{equation}
for a given charge. If
\begin{equation}
E_{\rm tot}(Q):={\rm min}_s\left[E^{(s)}_{\rm tot}(Q)\right]<0
\label{eq:Stable}
\end{equation}
a stable configuration is constructed. Figure ~\ref{fig_etot1} shows 
$E_{\rm tot}$ as a function of charge for various values of 
the Yukawa coupling constant, {\it i.e.} the mass of a 
non--interacting fermion.
For $f\approx 1.6$ the classical and fermion energies essentially cancel 
each other and leave $E_{\rm tot}$ roughly charge independent\footnote{The
exhibited dependence at small $Q$ is artificial because very narrow strings
have not been considered to avoid the Landau ghost inconsistency.}. Bound
objects are observed by further increasing the Yukawa coupling to about 
$f \approx 1.7$, which corresponds to a heavy fermion mass which is less than 
twice the top quark mass. We find that the minimizing configurations 
have $\xi_1\approx0$, {\it i.e.} they are dominated by the Higgs field.

\section{Conclusion}

We have presented and discussed the formalism to compute the fermion 
contribution to the vacuum polarization energy per unit length of an 
infinitely long straight string in a simplified version of the electroweak 
standard model. Our approach is based on the interface formalism, for 
which the analytical properties of scattering data are essential. We have 
also seen that a particular subset of gauge choices circumvents obstacles 
that in a na\"{\i}ve treatment arise from the non--trivial structure of 
the string configuration at spatial infinity. Numerically we have found 
that the vacuum polarization is small and positive in the regime in which 
the one--fermion loop approximation is reliable. Hence, there is no quantum 
stabilization of the string. However, we have seen that a heavy fermion 
doublet can stabilize a nontrivial string background for a non--zero fixed 
charge per unit length. The resulting configuration is dominated by the Higgs 
field.  Since any additional variational degree of freedom can only lower the 
total energy, the embedding of this configuration  in the full standard model, 
with the $U(1)$ gauge field included, will also yield a bound object as long 
as mixing between this heavy and the standard model fermions can be ignored. 
We see binding set in at $m \approx 300\,\mathrm{GeV}$, which is still within 
the range of energy scales at which the standard model is expected to provide 
an effective description of the relevant physics, and also within the 
range to be probed at the LHC. Light fermions would contribute only weakly to 
the binding of the string, since their Yukawa couplings are small. As a result,
we can add them to our model, {\it e.g.} to accommodate anomaly cancellation,
without significantly changing the result.

\section*{Acknowledgments}

One of us (HW) is grateful to the organizers for putting together this 
enjoyable conference and providing the opportunity to present this project.

This work is founded in parts by the NRF (South Africa). NG acknowledges 
support under NSF grant PHY08--55426.


\newpage

\end{document}